\documentclass[fleqn,12pt,twoside]{article}
\usepackage{espcrc1}
\usepackage{amssymb}

\usepackage{graphicx}
\usepackage[figuresright]{rotating}

\usepackage{wrapfig}

\newcommand{\AmS}{{\protect\the\textfont2
  A\kern-.1667em\lower.5ex\hbox{M}\kern-.125emS}}

\title{Two body scattering length of Yukawa model on a lattice}

\author{F. De Soto\address[upo]{D. Sistemas F\'{\i}sicos Qu\'{\i}micos y
Naturales. U. Pablo de Olavide, 41013 Sevilla
(Spain)}\address[lpsc]{Laboratoire de Physique Subatomique et
Cosmologie, 
38026 Grenoble (France)}, J.
Carbonell\addressmark[lpsc], C. Roiesnel\address[x]{Centre de
Physique Th\'eorique, Ecole
Polytechnique, 
91128 Palaiseau, (France)}, Ph.
Boucaud\address[lpt]{Laboratoire de Physique Th\'eorique,
Universit\'e Paris-XI, 91405 Orsay (France)}, J.P.
Leroy\addressmark[lpt] and O. P\`ene\addressmark[lpt]}

\begin{document}

\maketitle

\begin{abstract}
The extraction of scattering parameters from Euclidean simulations
of a Yukawa model in a finite volume with periodic boundary
conditions is analyzed both in non relativistic quantum mechanics
and in quantum field theory.
\end{abstract}

\section{Introduction}

A rigorous treatment of nuclei based on first principles quantum
chromodynamics requires the understanding of the transition from
the confining theory of quarks to the unconfined hadrons, a
present challenge for the high energy community. Only very
recently, the development of algorithms and computers devoted to
lattice QCD has made possible starting the study of
nucleon-nucleon~\cite{beane} systems (as well as $\pi\pi$, $KK$,
$\pi N$, etc) based on full QCD calculations. Nevertheless,
lattice methods have a drawback for computing scattering
parameters, as they can not be extracted in the infinite volume
limit~\cite{nogo}. However they can be obtained by measuring the
finite volume effects on the spectrum of a periodic boundary
condition box~\cite{Luscher}.

In this work, the scattering length of the Yukawa model has been
determined; on one hand, for the Yukawa potential $V(r)=-\alpha
e^{-m_sr}/r$ in Schrodinger equation, that can be precisely solved
in a 3-d torus, as well as in the infinite volume, serving as a
crosscheck for determining scattering length via finite volume
effects. On the other hand, the quantum field theory of fermions
($\psi$) and mesons ($\phi$) interacting via the lagrangian
$\mathcal{L}_I(x)=g_0\overline\psi(x)\Gamma\phi(x)\psi(x)$ -- that
gives rise to the latter potential (with $\alpha=g_0^2/4\pi$) in
the NR limit -- is solved in a Euclidean lattice.

The goals of this calculation are manyfold. First, the comparison
between the QFT and the non relativistic (NR) results can help to
disentangle the bias introduced by the ladder approximation, as
well as the size of the relativistic corrections. Second, the
methods used to extract the scattering length can be tested with
this model, much simpler than QCD. And finally, the interaction
itself is interesting, as scalar meson exchange is a common
ingredient of NN interaction potentials~\cite{OBEM}, responsible
for most of the nuclear binding energy.

\section{From bound state spectrum to low energy parameters}
\label{lep}

The relation between the spectrum of a system enclosed in a torus
and the phase-shifts in the infinite volume has been established
by M. Luscher~\cite{Luscher}. For a lattice with $L$ points in
each spacial direction and spacing $a$, the energies are
determined by the expression:
\begin{equation}\label{Luscher_S}
p \cot(\delta(p)) = \frac{1}{\pi aL}
S\left(\left(\frac{aLp}{2\pi}\right)^2\right)\ ,\qquad
S(\eta)=\lim_{\Lambda\to\infty} \sum_{\vec{n}\in
Z^3}^{|\vec{n}|\leq\Lambda}\frac{1}{n^2-\eta}-4\pi\Lambda\ .
\end{equation}
that is exact for lattice sizes $aL>2R$ where $R$ is the
interaction range, i.e. $V(x>R)=0$. For interactions of physical
interest this regime is reached exponentially and independently of
the coupling constant and the low energy parameters (LEP)
appearing in the effective range expansion, $p \cot(\delta(p))
\approx -\frac{1}{a_0} + \frac 1 2 r_0 p^2 + \cdots$. This
constitutes a remarkable advantage with respect to the, most
commongly used, large-L expansion of (\ref{Luscher_S}):
\begin{equation}\label{Luscher_A}
E \approx \frac{4\pi
a_0}{M(La)^3}\left(1+c_1\frac{a_0}{aL}+c_2\left(\frac{a_0}{aL}\right)^2+\cdots
\right)\ .
\end{equation}
\begin{wrapfigure}{R}{7.6cm} \vspace*{-1.5cm}
\includegraphics[width=7.4cm]{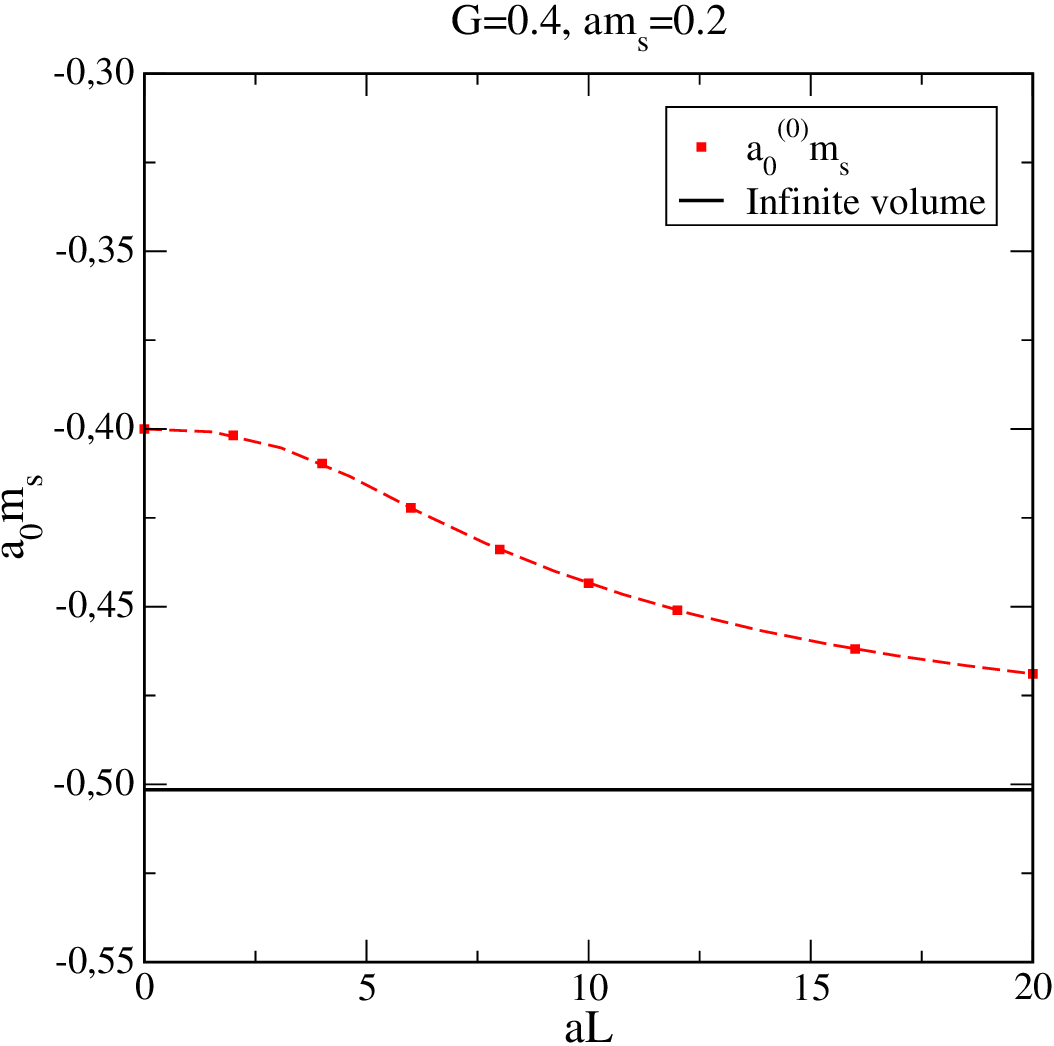}
\caption{\it \small Lattice result for $a^{(0)}_0$ vs volume
compared to the infinite volume result.} \label{a00}
\vspace*{-0.5cm}
\end{wrapfigure}

In a recent work \cite{NRmodel}, the possibility to extract the
LEP from the bound state spectra in a finite box has been examined
in the framework of the NR Yukawa model. The $L$-dependence of the
eigenenergies $E(L)$ in a 3d torus 
was used to extract the infinite volume scattering length $a_0$
and effective range $r_0$. 
Contrary to the quantum field lattice calculation, those
quantities can be independently computed by solving the
corresponding Schrodinger radial equation and used to test the
validity of the different approaches discussed in the literature.

The results were analyzed in terms of the slowly varying quantity
$a_0^{(0)}(L)\equiv (aL)^3 E(L)M/4\pi$ which represents the
zeroth-order approximation of large-L expansion (\ref{Luscher_A}).
The NR model depends on the unique parameter $G=\alpha M/m_s$, and
it was shown that $a_0^{(0)}(L)m_s$ tends to $-G$ in the small-L
limit, and -- according to (\ref{Luscher_A}) -- to the required
scattering length value $a_0$ for $aL\to\infty$ (figure
\ref{a00}). It is interesting to remark that $a_0^{(0)}(L)$
displays the same $1/L^3$ behavior in both limits.

\begin{wrapfigure}{R}{7.6cm}\vspace*{-1.5cm}
\includegraphics[width=7.4cm]{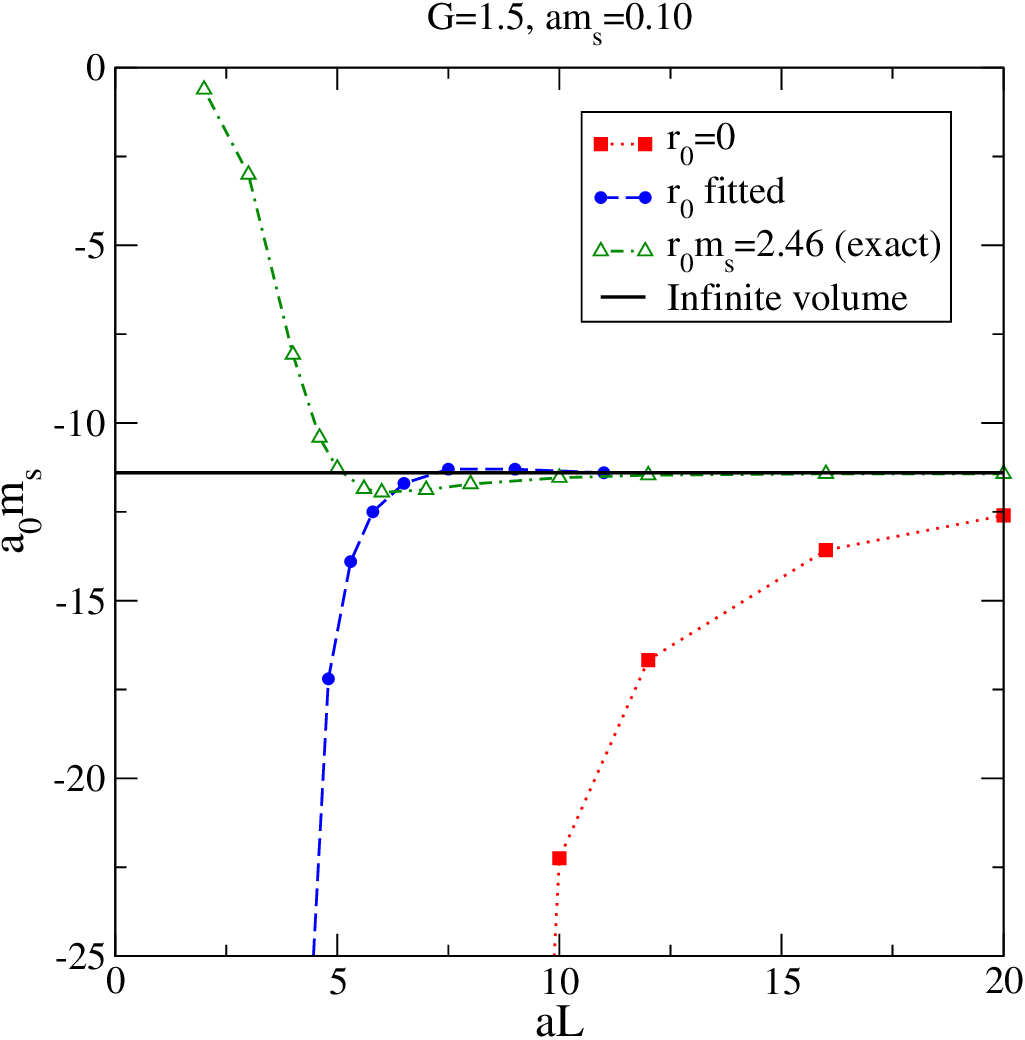}
\caption{\it \small $a_0$ obtained with equation (\ref{Luscher_S})
for different values of $r_0$. The influence of this parameter is
remarkable.} \label{a0S}\vspace*{-01cm}
\end{wrapfigure}

It was in particular found that one can obtain accurate values of
$a_0$ and $r_0$ by using equation (\ref{Luscher_S}) at two
different lattice sizes in the region $Lam_s\approx5$ and solving
the resulting linear system. Some results are illustrated in
figure \ref{a0S}. They show the sensibility to the effective range
$r_0$  and the need for both parameters to be simultaneously
determined.

The possibility of extracting LEP that way is independent of the
$a_0$ value and applies in the resonant case as well. Care must be
taken however when applying equation (\ref{Luscher_A}) to extract
the value of $a_0$. When using lattice sizes $Lam_s\sim 2-3$,
$E(L)$ is well fitted by a $c/L^3$ dependence but the coefficient
$c$ can strongly differ from the infinite volume scattering length
\cite{NRmodel}. Only the use of equation (\ref{Luscher_S}) at
lattice sizes greater than the interaction range --
$Lam_s\gtrsim5$ in the Yukawa model -- could lead to unambiguous
extraction of LEP, provided both $a_0$ and $r_0$ are taken into
account.

\section{Lattice results}

The non perturbative solutions of the Yukawa field theory in the
lattice have been considered in \cite{qcd05,qnp06}. The existence
of a critical value of the coupling has been
argued~\cite{qnp06,lee} hindering the appearance of bound states
of the QFT. The effect of the ladder approximation can
nevertheless be studied by calculating the scattering length and
comparing with the non relativistic results.

We have used Wilson discretization of fermion fields, which has
the disadvantage that there are $\mathcal{O}(a)$ discretization
errors. They can be important in our case, since nucleon mass is
not small. As the interaction is non confining, to reduce the
discretization errors we will use a hint from the free case, where
the time evolution of fermion is given by
\begin{equation}
C(t) = \sum_{\vec{x}}
Tr\langle\overline\psi(\vec{x},t)\psi(\vec{0},0)\rangle\ \propto
(1+aM_L)^{-t} \sim\ e^{-M t}\ .
\end{equation}
This expression defines the mass that governs the time evolution
of the nucleon $M$, and relates it to the fermion mass appearing
in the lagrangian, $M_L$, as $aM_L=\exp(aM)-1$.

Both masses are equivalent at the continuum limit, but they can
differ sizably when working at finite lattice spacing, what may
spoil the convergence of the lattice results. To overcome this, we
define an improved coupling, $\tilde{g}$, as the variation of
nucleon energy with the meson field at zero meson field (note that
the coupling acts as a mass term),
\begin{equation}\label{def_g}
\tilde{g}=\left.\frac{\partial E_N(\vec{p}=0)}{\partial
\phi^b}\right|_{\phi^b=0}\ .
\end{equation}
If we assume that the nucleon propagates in a small constant
background field, $\phi_b$, its mass is modified by the coupling
term as $aM_L\to aM_L + g_0\phi_b$ which implies $aM\to aM +
g_0\phi_b/(1+aM_L)$. The renormalized coupling defined above
results then
\begin{equation}
\tilde{g} = \frac{g_0}{1+aM}\ .
\end{equation}
Note that $g_0$ and $\tilde g$  are equivalent in the continuum
limit ($a\to 0$), but the second should have a better behavior at
finite lattice spacing. The usefulness of this coupling is
specially important when working with coarse lattices, as is the
case in nuclear physics, due to the mass hierarchy. Note that the
differences due to this definition of the coupling can be very
important, $\sim70\%$ for fermion masses $aM\sim0.3$.
\begin{wrapfigure}{R}{7.6cm}\vspace*{-.8cm}
\includegraphics[width=7.4cm]{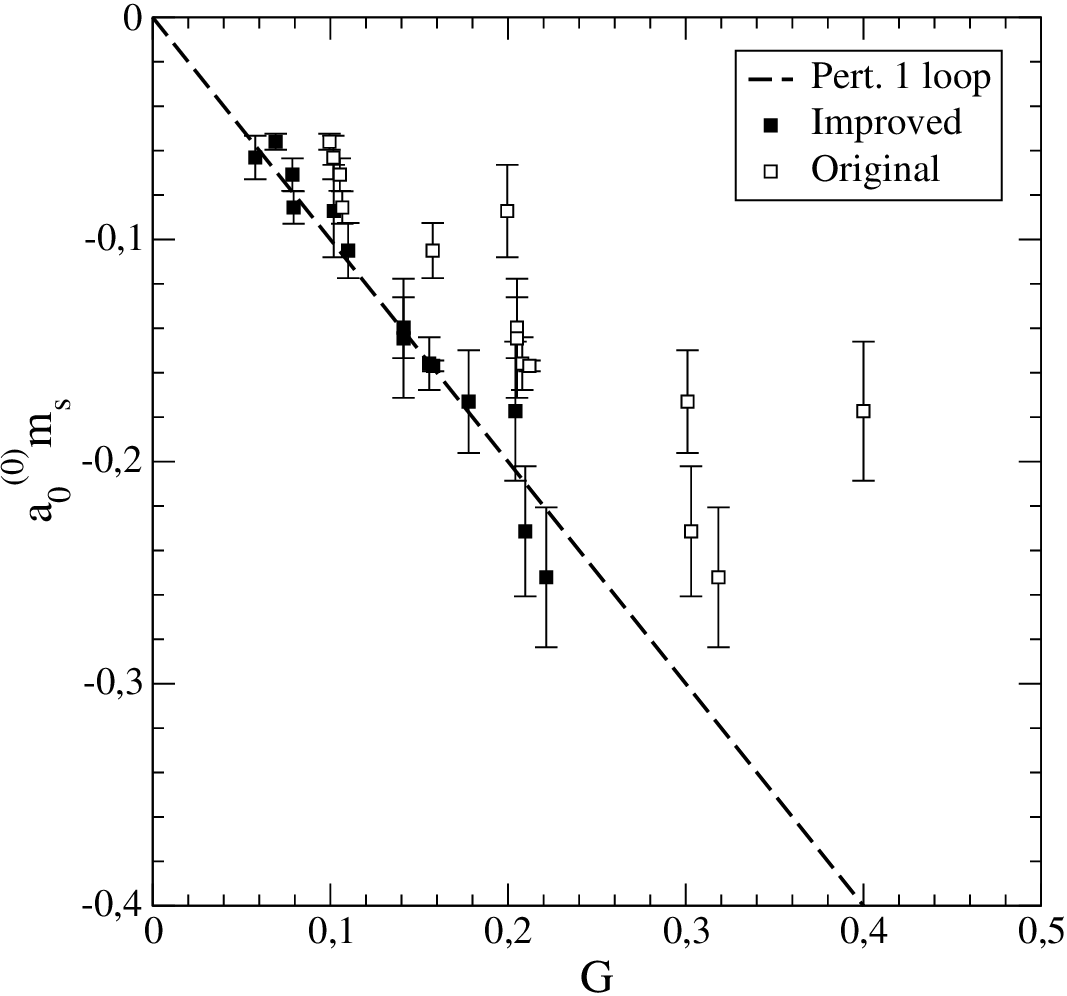}
\caption{Scattering length versus $G$ for the original $g_0$ and
modified coupling $\tilde g$ for lattice volumes $ Lam_s\sim
2-4$.} \label{a0lattice}\vspace*{-0.75cm}
\end{wrapfigure}

Two body S-wave scattering length can be computed from the
corresponding spectrum on a finite size lattice calculation as it
has been shown in the NR potential model. The fist test to be done
is the perturbative regime that, as in the NR case, predicts for
the scattering length $a_0 m_s = -G + \mathcal{O}(G^2)$. The
results of the scattering length obtained from lattice
calculations are shown in figure \ref{a0lattice} as a function of
the coupling constant $\alpha=g^2/4\pi$, both for the original and
the improved coupling. The correspondence with the small coupling
limit (dashed line) gets clearly better with the use of the
improvement factor (\ref{def_g}).

The fact that we do not see differences with the first order in
perturbation theory is maybe due to the fact that we work with
rather reduced lattice volumes. Further calculations have to be
done (in particular bigger volumes with $La\gtrsim 5m_s^{-1}$) in
order to compute the scattering length -- and effective range --
properly.

As a concluding remark, we have proposed an improved definition of
Yukawa coupling constant in the lattice that exhibits quite better
scaling properties than the bare one. As shown in figure
\ref{a0lattice}, this correction factor is mandatory to test
perturbation theory -- and accordingly to obtain any physical
result beyond this regime -- even for rather small lattice
spacings $aM\sim0.2$.

\end{document}